\documentclass[preprint2,graphics]{aastex}

\slugcomment{to appear in Astrophysical Journal}

\shorttitle{Type Ia supernovae enrichment in $\omega$~Cen?}
\shortauthors{Pancino et al.}

\begin{document}


\title{High resolution spectroscopy of metal rich giants in $\omega$
Cen:\\first indication of SNe~Ia enrichment\thanks{Based on UVES
observations collected at the European Southern Observatory, Paranal,
Chile, within the Observing Programme 165.L-0263. Also based on WFI
observations collected at La Silla, Chile, within the observing
programmes 62.L-0354 and 64.L-0439.}}


\author{E. Pancino\altaffilmark{1}, L. Pasquini}
\affil{European Southern Observatory, K. Schwarzschild Str. 2, 
Garching, D-85748, Germany.}
\email{epancino@eso.org, lpasquin@eso.org}

\author{V. Hill} 
\affil{Observatoire de Paris-Meudon, F-92195 Meudon, France}
\email{Vanessa.Hill@obspm.fr}

\and

\author{F. R. Ferraro and M. Bellazzini}
\affil{Osservatorio Astronomico di Bologna, Via Ranzani 1, I-40127, Bologna, 
Italy.}
\email{ferraro@apache.bo.astro.it, bellazzini@bo.astro.it}


\altaffiltext{1}{on leave from Dipartimento di Astronomia,
Universit\`a di Bologna, Via Ranzani 1, I-40127, Bologna, Italy.}

\begin{abstract}

We have obtained high-resolution, high S/N spectra for six red giants
in $\omega$~Cen: three belonging to the recently discovered,
metal-rich Red Giant Branch \citep[RGB-a, as defined in][]{p00} and
three to the metal intermediate population (RGB-MInt). Accurate Iron,
Copper and $\alpha$-elements (Ca and Si) abundances have been derived
and discussed. In particular, we have obtained the first direct
abundance determination based on high-resolution spectroscopy for the
RGB-a population, $<$[Fe/H]$>=-0.60\pm 0.15$. Although this value is
lower than previous estimates based on Calcium triplet measurements,
we confirm that this population is the most metal rich in
$\omega$~Cen. In addition, we have found a significant difference in
the $\alpha$-elements enhancement of the two populations. The three
RGB-MInt stars have the expected overabundance, typical of halo and
globular clusters stars: $<$[$\alpha$/Fe]$>=+0.29\pm 0.01$. The three
RGB-a stars show, instead, a significantly lower $\alpha$-enhancement:
$<$[$\alpha$/Fe]$>=+0.10\pm 0.04$. We have also detected an increasing
trend of [Cu/Fe] with metallicity, similar to the one observed for
field stars by \citet{sneden}. The observational facts presented in
this letter, if confirmed by larger samples of giants, are the first
indication that supernovae type Ia ejecta have contaminated the medium
from which the metal rich RGB-a stars have formed. The implications
for current scenarios on the formation and evolution of $\omega$~Cen
are briefly discussed.
 
\end{abstract}



\section{Introduction}
\label{intro}

\begin{deluxetable}{lrrrrr}
\tabletypesize{\footnotesize}
\tablecaption{Literature Data for Star ROA~371. \label{371}}
\tablehead{
\colhead{ } & \colhead{PN89} & \colhead{B91} &
\colhead{V94} & \colhead{ND95} & \colhead{Here}
} 
\startdata
$R$       &$\sim17000$&$\sim17000$&$\sim20000$&$\sim38000$&$\sim45000$\\
$S/N$     &$\leq 50$  &$\sim 100$ & 70--150   &$\sim 50$  & 140       \\
$T_{eff}$ & 4000      & 4000      & 4000      & 4000      & 4000      \\
$\log g$  & 0.9       & 0.9       & 0.9       & 0.9       & 0.7       \\
$v_t$     & 2.5       & 1.5       & 2.2       & 1.6       & 1.5       \\
$$[Fe/H]  & -1.37     & -0.9      & -1.00     & -0.79     & -0.95     \\
\enddata
\tablecomments{Different literature results for star ROA~371 are
summarized in each column: {\em PN89} from \citet{p&n}; {\em B91} from
\citet{brown}; {\em V94} from \citet{vanture}; {\em ND95} from
\citet{norris} and {\em Here} from this Letter.}
\end{deluxetable}

The globular cluster $\omega$~Cen is the most massive and luminous of
all galactic globular clusters, and it is the {\it only} one which
shows undisputed variations in its heavy elements content. For these
reasons, $\omega$~Cen red giants have been the subject of the largest
spectroscopic surveys ever attempted on a globular cluster. Calcium
triplet, low-resolution studies \citep{norris96,s&k} have shown that
{\em (i)} few stars exist on the Red Giant Branch (RGB) with
[Fe/H]$<-1.8$; {\em (ii)} there is a well defined peak in the
distribution at [Fe/H]$=-1.6$ with a long, extended tail reaching
[Fe/H]$=-0.5$; {\em (iii)} the distribution appears bimodal with a
second, smaller peak at [Fe/H]$\sim-1.0$.

In spite of the previous massive observational efforts, the full
complexity of the RGB structure of $\omega$~Cen has been revealed only
recently, thanks to wide field photometry \citep{lee,p00} that
discovered the presence of a previously undetected, anomalous sequence
(hereafter RGB-a) on the red side of the main RGB. This newly
discovered population is thought to represent the rich end of the
metallicity distribution in $\omega$~Cen: \citet{p00} estimated a mean
[Ca/H]$\simeq-0.1$ and [Fe/H]$\simeq-0.4$ from six RGB-a stars in
common with previous surveys \citep[stars ROA~300, 447, 500,
513\footnote{in \citet{p00}, this star was erroneously reported as
star ROA~512.}, 517 and 523; nomenclature from][]{woolley}.

In the framework of a coordinated spectro--photometric project devoted
to the study of this puzzling stellar system, we present here the
first high resolution abundance measurements of the most metal rich
giants in $\omega$~Cen.

\section{Observational material}

\begin{deluxetable}{llccccccccc}
\tabletypesize{\footnotesize}
\tablecaption{Programme stars. \label{param}}
\tablehead{\colhead{ROA} & \colhead{WFI} & \colhead{$V$} & 
\colhead{$M_V$} & \colhead{$(B$-$V)_0$} & \colhead{$T_{e}(B$-$V)$} & 
\colhead{$\log g (M_V)$} & \colhead{$T_{e}$(Fe)} & 
\colhead{$\log g$(Fe)} & \colhead{$v_t$} & \colhead{$\rm[Fe/H]$}}
\startdata
300 & 221132 & 12.71 & -1.21 & 1.48 & 3800 & 0.7 & 3900 & 0.7 & 1.4 & -0.77 \\
--- & 222068 & 12.95 & -0.97 & 1.42 & 3900 & 0.9 & 4000 & 1.1 & 1.3 & -0.49 \\
--- & 222679 & 13.26 & -0.66 & 1.26 & 4100 & 1.2 & 4100 & 1.2 & 1.4 & -0.54 \\
211 & 619210 & 12.43 & -1.49 & 1.37 & 3950 & 0.8 & 4000 & 0.8 & 1.9 & -1.02 \\
371 & 617829 & 12.71 & -1.21 & 1.34 & 4000 & 0.9 & 4000 & 0.7 & 1.5 & -0.95 \\
--- & 618854 & 13.26 & -0.66 & 1.00 & 4600 & 1.6 & 4600 & 1.2 & 1.5 & -1.20 \\
\enddata
\tablecomments{The table columns contain the following information:
{\em(1)} the Royal Astronomical Observatory (ROA) number from
\citet{woolley}; {\em(2)} the WFI catalog number from \citet{p00};
{\em(3)-(5)} $V$ magnitude, $M_V$ absolute magnitude and $(B-V)_0$
dereddened color (see text); {\em (6)-(7)} photometric estimates of
$T_{eff}$ and $\log g$ based on \citet{monte}; {\em (8)-(11)} stellar
parameters $T_{eff}$, $\log g$, $v_t$ and [Fe/H] derived from our
abundance analysis (see text).}
\end{deluxetable}

We selected our targets among the metal rich giants in
$\omega$~Cen. Three of them, stars ROA~300, WFI~222068 and WFI~222679,
belong to the RGB-a, while the other three belong to the intermediate
metallicity population \citep[RGB-MInt, defined in][]{p00}. Only star
ROA~371 (Table~\ref{371}) has been observed before with
high-resolution spectroscopy ($R\geq20000$) by \citet{p&n},
\citet{brown}, \citet{vanture} and \citet{norris}, and for this reason
is the ideal comparison object.

Observations were carried out in June 2000 with UVES at the ESO Very
Large Telescope (VLT) Kueyen at Paranal, as a backup programme while
the main targets were not visible. We obtained high-resolution ($R\sim
45000$) echelle spectra with $S/N\sim 100-150$ per resolution
element. The monodimensional spectra were extracted with the UVES
pipeline \citep{uves}, then continuum-normalized and corrected for
telluric absorption bands with IRAF\footnote{IRAF is distributed by
the National Optical Astonomy Observatories, which is operated by the
association of Universities for Research in Astronomy, Inc., under
contract with the National Science Foundation.}. Radial velocities
measured on our spectra confirm membership for all the six stars.

The magnitudes and colors of our programme stars (see
Table~\ref{param}), used to estimate $T_{eff}$ and $\log g$, are from
\citet{p00} and from unpublished V data, obtained during the same
observing runs and treated the same way. Dereddened colors and
absolute magnitudes were derived, assuming $E(B-V)=0.12$ and
$(m-M)_V=13.92$ \citep{harris}. Effective temperatures and surface
gravities were obtained with the calibration by \citet{monte} for
giants, assuming a mass of $0.7~M_{\odot}$.

\section{Abundance Analysis}

\begin{deluxetable}{llllll}
\tabletypesize{\footnotesize}
\tablecaption{Resulting Element Abundances. \label{elements}}
\tablehead{\colhead{Star} & \colhead{[Fe/H]} & \colhead{[Ca/Fe]} &
\colhead{[Si/Fe]} & \colhead{[Cu/Fe]} & \colhead{Pop.}} 
\startdata
ROA~300    & -0.77$\pm$0.02 ($\pm$0.10) 
           & +0.12$\pm$0.07 ($\pm$0.15)  
           & +0.01$\pm$0.11 ($\pm$0.11)
           & -0.32$\pm$0.13 ($\pm$0.04)
	   & RGB-a\\
WFI~222068 & -0.49$\pm$0.02 ($\pm$0.12)
           & +0.15$\pm$0.06 ($\pm$0.16)
           & +0.11$\pm$0.10 ($\pm$0.11)
           & -0.15$\pm$0.12 ($\pm$0.05)
           & RGB-a\\ 
WFI~222679 & -0.54$\pm$0.02 ($\pm$0.11)
           & +0.06$\pm$0.05 ($\pm$0.14)
           & +0.08$\pm$0.10 ($\pm$0.11)
           & -0.33$\pm$0.11 ($\pm$0.08)
           & RGB-a\\
ROA~211    & -1.02$\pm$0.01 ($\pm$0.08)
           & +0.28$\pm$0.04 ($\pm$0.14)
           & +0.28$\pm$0.07 ($\pm$0.10)
           & -0.45$\pm$0.09 ($\pm$0.12)
           & RGB-Int\\
ROA~371    & -0.95$\pm$0.01 ($\pm$0.09)
           & +0.30$\pm$0.03 ($\pm$0.14)
           & +0.26$\pm$0.11 ($\pm$0.11)
           & -0.33$\pm$0.07 ($\pm$0.04)
           & RGB-Int\\
WFI~618854 & -1.20$\pm$0.01 ($\pm$0.12)
           & +0.28$\pm$0.04 ($\pm$0.05)
           & +0.30$\pm$0.06 ($\pm$0.10)
           & -0.41$\pm$0.04 ($\pm$0.07)
           & RGB-Int\\
\enddata
\end{deluxetable}

We selected a set of reliable and unblended spectral lines that span a
wide range in strength, excitation potential and wavelength. In
particular, the results presented here are based on 94
\ion{Fe}{1} lines, 10 \ion{Fe}{2} lines, 17 \ion{Ca}{1} lines, 10
\ion{Si}{1} lines and the $5782$\AA\ line for \ion{Cu}{1}, all
in the spectral range $\sim 5300 - 6800$\AA. Atomic data were taken
mainly from the NIST\footnote{NIST (National Institute of Standards
and Technology) Atomic Spectra Database, Version 2.0 (March 22, 1999),
\tt {http://physics.nist.gov/cgi-bin/AtData/main\_asd}.}  database,
and from \citet{nave} for Iron. For Copper we used the hyperfine
structure line list from the Kurucz database
\citep{bielski}.

Equivalent widths (EW) for Fe, Ca and Si were measured with IRAF by
gaussian fitting of the line profile on the local continuum. The
gaussian profile is a good approximation for these stars, if one
avoids strong lines. For the coolest stars we chose {\it not} to
measure any atomic line inside the prominent TiO bands. The comparison
of our EW measurements for ROA~371 with \citet{norris-data} shows a
modest ($\sim2\%$) systematic difference that has a negligible impact
on the final abundance determination: the results for ROA~371 are, in
fact, in good agreement with other literature values
(Table~\ref{371}).

The abundance calculations were made using an extension of the OSMARCS
grid of plane parallel, LTE model atmospheres for M giants and
supergiants calculated by \citet{plez} and Plez (1995, private
communication). First estimates of the microturbulent velocity $v_t$
and of [Fe/H] were derived from curves of growth for \ion{Fe}{1} and
\ion{Fe}{2}. We then explored the parameters space around our first
estimates, and refined them by enforcing simultaneously the following
conditions: {\it (a)} $v_t$ was determined by imposing that strong and
weak \ion{Fe}{1} lines give the same abundance; {\it (b)} the
temperature was constrained by imposing excitation equilibrium on
\ion{Fe}{1}; {\it (c)} the surface gravity was refined by imposing
ionization equilibrium between \ion{Fe}{1} and \ion{Fe}{2}. As a final
test, we checked that [Fe/H] was not changing with $\lambda$. The
resulting best parameters for the six programme stars are shown in
Table~\ref{param}: temperatures and gravities are close to the
photometric estimates, and we adopted them to derive the abundances of
the other elements (Ca and Si). For Copper we computed the line
profile with a spectral synthesis technique to take into account the
hyperfine structure of the 5782\AA\ line \citep{bielski}.

Table~\ref{elements} shows the final abundances, together with their
uncertainties: the first figure quoted is the random uncertainty,
obtained by averaging the abundances from single lines. In the case of
Copper, the error reflects the goodness of the spectral synthesis
fit. The uncertainty in parenthesis is instead an estimate of the
internal systematics, obtained by altering the values of $T_{eff}$ (by
$\pm$100~$K$), $\log g$ (by $\pm$0.2) and $v_t$ (by
$\pm$0.2~$km~s^{-1}$) and by evaluating the impact of such changes on
the resulting abundances. To give an idea of the (external) systematic
uncertainties, we analyzed the spectrum of Arcturus ($\alpha$~Boo),
taken with UCLES at the Anglo Australian Telescope (Bessel, private
communication), with the same procedure used for the programme stars,
obtaining: $T_{eff}=4300~K$, $\log g=1.6$, $v_t=1.4~km~s^{-1}$ and
[Fe/H]$=-0.43$. The reader can use this, together with the abundance
analisys of ROA~371 (Table~\ref{371}) to place our results in a more
general context.

\section{Results}

The three RGB-a stars are the first members of this population ever
analyzed with high-resolution spectroscopy. By taking a straight
average of their abundances (from Table~\ref{elements}) we found
[Fe/H]=$-0.60\pm 0.15$ and [Ca/H]=$-0.49\pm 0.16$, confirming that the
RGB-a population is the most metal rich component of the $\omega$~Cen
stellar mix. The RGB-a abundance obtained here is lower than the
previous estimate ([Ca/H]$\sim -0.1$) based on Calcium triplet surveys
(see Par.~\ref{intro}). However, Calcium triplet calibrations are
usually rather uncertain in the high metallicity regime
\citep[see][]{norris96}.
 
Our most interesting and surprising result, however, is the different
$\alpha$-elements enhancement found for the two sub-populations. We
obtained [Ca/Fe]=$+0.29\pm 0.01$ and [Si/Fe]=$+0.28\pm 0.02$ for the
three RGB-MInt stars, which is the expected value for the
$\alpha$-enhancement of halo and globular cluster stars. Instead, the
three RGB-a stars have only [Ca/Fe]=$+0.11\pm 0.05$ and
[Si/Fe]=$+0.07\pm 0.05$. If we compute [$\alpha$/Fe] for each star
with a weighted average of their Ca and Si abundances, the RGB-MInt
and RGB-a populations $\alpha$-enhancements turn out to be
[$\alpha$/Fe]$=+0.29\pm 0.01$ and [$\alpha$/Fe]$=+0.10\pm 0.04$,
respectively.

This is the first indication that a population of stars with a
significantly lower $\alpha$-enhancement exists in $\omega$~Cen: both
\citet{norris} and \citet{smith}, who analyzed giants in $\omega$ Cen with
metallicities up to [Fe/H]$=-0.78$ and [Fe/H]$=-0.95$ respectively,
found {\em no evidence} of a decrease in Calcium, Silicon or in any
other $\alpha$-element enhancement.

\begin{figure}[tbh]
\epsscale{1.0}
\plotone{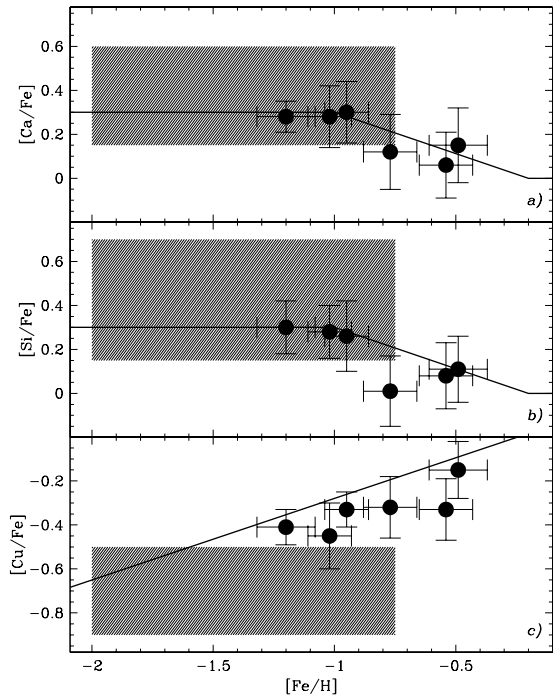}
\caption{Element ratios for [Ca/Fe] ({\it a)}), [Si/Fe] ({\it b)}) and
[Cu/Fe] ({\it c)}). Our results are plotted as filled dots, while the
shaded areas mark the region where previous, high resolution
measurements for $\omega$~Cen giants lie and the solid lines represent
the corresponding trends in our Galaxy (see text for
references). \label{cafe}}
\end{figure}

The effect is illustrated in Figure~\ref{cafe} (Panels {\it a)} and
{\it b)}), where the Calcium and Silicon enhancements are plotted as a
function of [Fe/H]. The shaded areas mark the region where the
measurements by \citet{norris} and \citet{smith} lie, while the solid
lines represents the Galactic [$\alpha$/Fe] relation as can be derived
from \citet{edvardsson} and \citet{gratton}. The six stars analyzed
here are plotted as filled dots: as can be seen, the three RGB-MInt
stars show enhancements that are in good agreement with previous
determinations for $\omega$~Cen, while the three RGB-a stars are
clearly less overabundant in both elements. We would like to note at
this point that even if some small systematic differences may be
present among the various studies (most probably due to a different
choice of atomic data and $\log gf$, especially in the case of Copper)
the use of trends in the abundance ratios is particularly reliable
since it is based on relative, differential measurements.

Panel {\it c)} of Figure~\ref{cafe} shows our results for [Cu/Fe]. Our
stars are plotted as filled dots, the shaded area covers the region
where \citet{smith} measurements lie, and the solid line represents
the trend found by \citet{sneden} for field stars. The increasing
trend in our data is more compatible with the results by
\citet{sneden} than with the behaviour of the metal poor stars in
$\omega$~Cen. As discussed by \citet{smith}, Copper is thought to be
produced mainly by type Ia supernovae, and only marginally by type II
supernovae \citep[see also][]{matteucci93}, as required to explain the
trend observed in the Galaxy. For the metal poor and intermediate
stars in $\omega$~Cen, \citet{smith} found a very low and constant
value of [Cu/Fe]=$-0.6$, in agreement with the idea that SNe~Ia have
{\it not} contributed to the enrichment of these two populations.

If the results shown in Figure~\ref{cafe} are confirmed by larger
samples of metal rich stars, then {\em we have found the first
evidence that Supernovae type Ia ejecta have contaminated the medium
from which the RGB-a stars have formed.}

\section{Discussion}

A detailed discussion on the chemical evolution of $\omega$~Cen is
beyond the purpose of this Letter, however some obvious implications
of the new results presented here deserve a short comment. There are
two main scenarios that have been proposed to explain the variety of
stellar populations observed in $\omega$~Cen, both based upon
non-negligible observational evidences: the {\em self-enrichment}
scenario and the {\em merging} scenario. Let us discuss the impact of
our result on the two separately, while recalling that the actual
evolutionary path of $\omega$~Cen may well be the result of a
combination of both scenarios.

\smallskip
{\em A) The self-enrichment scenario.} The most popular explanation is
that $\omega$~Cen has been somehow able to retain the ejecta of
previous generations of stars and to self-enrich during its star
formation history. This requires that $\omega$~Cen was formerly a more
complex and larger stellar system, with a deeper potential well,
possibly a dwarf galaxy \citep{freeman,dinescu,hk00,hughes}, that lost
most of its stars (and gas) in the interaction with the Milky Way, as
the Sagittarius dSph is doing presently.

The fundamental piece of evidence in favour of this scenario is the
constancy of the $\alpha$-enhancement with metallicity
([$\alpha$/Fe]$\sim$+0.3) for all stars with [Fe/H]$\le-0.8$
\citep{norris,smith}. This requires that the major responsibles for
the enrichment of these stars are type II supernovae: at least part of
their gas must have been retained by $\omega$~Cen in order to explain
the iron abundance spread. The timescales of this process are thought
to be relatively short ($\le$ 1 Gyr), in contradiction with the
timescales required to explain the observed dramatic increase of the
$s$-process elements overabundance with metallicity
\citep{norris,smith,evans}. The major responsibles of the $s$-process
elements enrichment are intermediate mass AGB stars, that act on a
timescale of at least $1-5~Gyr$ \citep{busso}\footnote{An alternative
explanation for the $s$-process enrichment is proposed by
\citet{ventura}, who examine the possibility of surface pollution for
stars in globular clusters.}. Thus, we are still far from assessing
robust timescales for the various enrichment processes in
$\omega$~Cen.

The $\alpha$ and $s$-process elements trends, together, have been
taken as evidence that no significant contribution by type Ia
supernovae have enriched the (metal poor and intermediate) stars in
$\omega$~Cen \citep{smith}. This has been explained either with a star
formation process short enough to end before the onset of type Ia
supernovae, or by assuming that type Ia supernovae winds efficiently
removed most of their own products \citep{recchi}. However, if we
assume that the detailed chemical composition of {\em all} giants in
$\omega$~Cen can be explained by pure self-enrichment, then the
evidence presented here indicates that $\omega$~Cen was able to retain
part of the SNe~Ia ejecta as well, complicating the picture.

If this is the case, then the RGB-a stars must be younger, possibly
the result of the last burst of star formation in
$\omega$~Cen. Indeed, an age spread of $3-5~Gyr$ has been claimed to
explain the morphology of the Sub Giant Branch-Turn Off
\citep{hk00,hughes}. In fact, we would have detected the ``knee'' of
the [$\alpha$/Fe] relation, a very valuable constraint to the chemical
evolution of the whole stellar system (McWilliam 1997). The canonical
interpretation of such feature is that it marks the onset of the
supernovae type Ia pollution. While the timescale of this enrichment
process are generally believed to be short ($\le$ 1 Gyr), they are
still quite uncertain, since they depend on several factors
\citep{matteucci}.

\smallskip
{\em B) The merging scenario.} The reason why this alternative
scenario \citep{searle,icke} is not ruled out yet, lies mostly in the
dynamical and structural properties of $\omega$~Cen. In particular,
the unusually high ellipticity of $\omega$~Cen, that has been
demonstrated to be sustained by rotation \citep{merritt}, is
compatible with the flattened shapes resulting from the merger of two
globular clusters \citep{makino}. Moreover, \citet{norris97} showed
that only stars with [Fe/H]$\le-1.2$ in $\omega$~Cen do rotate, while
the more metal rich ones show no evident sign of rotation. Finally,
\citet{p00} showed that, while the metal poor population exhibits the
well known East-West elongation, the two metal rich populations show a
more pronounced ellipticity, but with an elongation in the North-South
direction. These pieces of evidence point toward a different dynamical
origin for the various sub-populations in $\omega$~Cen.

There are some problems with this scenario, though. According to
\citet{norris96} and \citet{smith}, the simple merging of two or more
single metallicity clusters cannot account for the broad metallicity
distribution of the RGB. Also, the high speed of ordinary, already
formed globular clusters in the potential well of the Milky Way makes
this kind of merging quite unlikely. However, the possibility still
remains that some of the present stellar components may have formerly
belonged to an external, smaller system that merged with $\omega$~Cen.

Within this framework, one of the most promising ideas seems to be the
so-called {\em merger within a fragment scenario} \citep{norris97}.
According to it, the RGB-a population could originally have been a
satellite system which fell into the potential well of the larger
stellar system (a small galaxy) whose final remnant is the
$\omega$~Cen that we observe today. In this case, our results add the
evidence that the accreted sub-system has previously been enriched by
type Ia supernovae, as opposed to the main body of $\omega$~Cen.
Thus, the original environments from which the sub-populations have
formed must have had different chemical enrichment histories.

\acknowledgments

It is a pleasure to thank R. Gratton, J. E. Norris, L. Origlia,
F. Primas, M. Rejkuba and M. Zoccali for their help and for many
useful discussions. E. P. acknowledges the support of the ESO
Studentship Programme. The financial support of the ASI (Agenzia
Spaziale Italiana) and of the italian MURST (Ministero della
Universit\`a e della Ricerca Scientifica e Tecnologica) to the project
{\it ``Stellar Observables of Cosmological Relevance''} is kindly
acknowledged.

\end{document}